\documentclass[twocolumn]{aastex63}
\usepackage{amsmath,amssymb}
\usepackage{wasysym}
\usepackage{graphicx,epsf,xcolor}
\usepackage{ulem}

\begin{document}

\title{Velocity Dispersions of Massive  Quiescent Galaxies from Weak Lensing and Spectroscopy\footnote{Based on data collected at Subaru Telescope, which is operated by the National Astronomical Observatory of Japan.}}
\author[0000-0001-6161-8988]{Yousuke Utsumi}\email{youtsumi@slac.stanford.edu}
\affiliation{Kavli Institute for Particle Astrophysics and Cosmology, SLAC National Accelerator Laboratory, Stanford University, 2575 Sand Hill Road, Menlo Park, CA 94025, USA}
\author[0000-0002-9146-4876]{Margaret J. Geller}
\affiliation{Smithsonian Astrophysical Observatory, 60 Garden Street, Cambridge, MA 02138, USA}
\author[0000-0003-1672-8234]{Harus J. Zahid}
\affiliation{Smithsonian Astrophysical Observatory, 60 Garden Street, Cambridge, MA 02138, USA}
\author[0000-0002-9254-144X]{Jubee Sohn}
\affiliation{Smithsonian Astrophysical Observatory, 60 Garden Street, Cambridge, MA 02138, USA}
\author[0000-0003-0751-7312]{Ian P. Dell'Antonio}
\affiliation{Department of Physics, Brown University, Box 1843, Providence, RI 02912, USA}
\author{Satoshi Kawanomoto}
\affiliation{National Astronomical Observatory of Japan, 2-21-1 Osawa, Mitaka, Tokyo 181-8588, Japan}
\author[0000-0002-3852-6329]{Yutaka Komiyama}
\affiliation{National Astronomical Observatory of Japan, 2-21-1 Osawa, Mitaka, Tokyo 181-8588, Japan}
\author[0000-0002-9679-9376]{Shintaro Koshida}
\affiliation{Subaru Telescope, National Astronomical Observatory of Japan, 650 North A'ohoku Place Hilo, HI 96720, USA}
\author[0000-0002-1962-904X]{Satoshi Miyazaki}
\affiliation{National Astronomical Observatory of Japan, 2-21-1 Osawa, Mitaka, Tokyo 181-8588, Japan}

\begin{abstract}
We use MMT spectroscopy and deep Subaru Hyper Suprime-Cam (HSC) imaging to compare the spectroscopic central stellar velocity dispersion of quiescent galaxies 
with the effective dispersion of the dark matter halo derived from the stacked lensing signal. The spectroscopic survey (the Smithsonian Hectospec Lensing Survey) provides a sample of 4585 quiescent galaxy lenses with measured line-of-sight central stellar velocity dispersion ($\sigma_{\rm SHELS}$) that is more than 85\% complete for $R < 20.6$, $D_{n}4000> 1.5$ and $M_{\star} > 10^{9.5}{\rm M}_{\odot}$.
The median redshift of the sample of lenses is 0.32. We measure the stacked lensing signal from the HSC deep imaging. The central stellar velocity dispersion is directly proportional to the velocity dispersion derived from the lensing $\sigma_{\rm Lens}$, $\sigma_{\rm Lens} = (1.05\pm0.15)\sigma_{\rm SHELS}+(-21.17\pm35.19)$.
The independent spectroscopic and weak lensing velocity dispersions probe different scales, $\sim3$kpc and  $\gtrsim$ 100 kpc, respectively, and
strongly indicate that the observable central stellar velocity dispersion for quiescent galaxies is a good proxy for the velocity dispersion of the dark matter halo.
We thus demonstrate the power of combining high-quality imaging and spectroscopy to shed  light on the connection between galaxies and their dark matter halos.
\end{abstract}


\section{Introduction}

In the standard hierarchical structure formation scenario, each observable galaxy inhabits a dark matter halo. Probing the relationship between the
galaxy and its halo is a continuing challenge. Quiescent galaxies have been central to these investigations partly because they are the most massive galaxies. 

We focus on the central stellar velocity dispersion  of quiescent galaxies, a quantity that is well-correlated with other characteristic observables \citep{1976ApJ...204..668F,1987ApJ...313...59D}. The velocity dispersion provides a promising  connection between the observable galaxy with its dark matter halo \citep[e.g.][]{schechter2015new,2012ApJ...751L..44W,2013AA...549A...7V,2016ApJ...832..203Z}.

For a small sample of nearby quiescent objects, \citet{1976ApJ...204..668F} first demonstrated the clear correlation between the central stellar velocity dispersion and galaxy luminosity.
Since then, others have demonstrated this correlation for larger samples and they have identified scaling 
relations between the central stellar velocity dispersion and other observables \citep[e.g.][]{Bernardi2003,Bernardi2004,Spindler2017}.
The empirically determined fundamental plane relating the velocity dispersion, the effective radius and the mean surface brightness within the effective radius has been an important benchmark in the study of quiescent galaxies \citep{1987ApJ...313...59D}.
\citet{Bernardi2003,Hyde2008} recast the fundamental plane in terms of the stellar mass  accounting, in principle, for mass-to-light ratio variations.
\citet{2013MNRAS.432.1862C} further investigate the relationship between stellar mass and velocity dispersion. The observed relations provide the foundation for examining the relationship between the velocity dispersion and the dark matter halo as probed by lensing observations.

Weak gravitational lensing provides a powerful tool for connecting  observable parameters with the otherwise invisible dark matter halo (e.g., \citealp{2001PhR...340..291B}).
From the first detection of the weak lensing signal associated with an  ensemble of  galaxies \citep{1984ApJ...281L..59T}, a host of studies have provided  more and more impressive  measures of the dark matter profiles as a function of various observables \citep{1996ApJ...466..623B,1998ApJ...503..531H,2000AJ....120.1198F,2001astro.ph..8013M,2001ApJ...551..643S,2003MNRAS.340..609H,2004ApJ...606...67H,2004AJ....127.2544S,2006MNRAS.368..715M,2007ApJ...667..176G}.
\citet{2013AA...549A...7V} derive the amplitude of the weak lensing signal around galaxies as a function of the central stellar velocity dispersion to determine the projected distribution of dark matter. They find that the lensing signal of galaxies is equally well traced by the stellar mass and the central stellar velocity dispersion. Strong lensing has also provided important constraints on the properties of the dark matter halo. For example, \citet{2014ApJ...793...96S} show that ``velocity dispersions'' estimated from Einstein ring radii  result in a fundamental plane that is substantially tighter than the classical relation based on spectroscopically measured stellar velocity dispersions.

\citet{2012ApJ...751L..44W} and \citet{2015ApJ...800..124B} argue that the central stellar velocity dispersion of a quiescent galaxy is a probe of its host dark matter halo. 
\citet{schechter2015new} suggests that that the observed stellar velocity dispersion is a good proxy for the halo velocity dispersion.
\citet{Zahid_2018} use the Illustris-1 simulations \citep{Nelson2015} to show that the observed central stellar velocity dispersion is proportional to the dark matter halo velocity dispersion for both central and satellite galaxies; for satellite galaxies the central stellar velocity dispersion traces the dark matter halo velocity dispersion at the time of infall.

Here we combine deep Hyper Suprime-Cam (HSC, \citealp{Miyazaki2018}) imaging with MMT spectroscopy
\citep{2005PASP..117.1411F} to compare the weak lensing signal  with spectroscopic central stellar velocity dispersions for 4585 quiescent galaxies.
Our study uses the deeper HSC imaging to extend the similar analysis of \citet{2013AA...549A...7V} to greater redshift.
We also compare the observational results with predictions based on the Illustris-1 simulations.

We describe the imaging and spectroscopy in Section \ref{sec:data}.
We outline the method of analysis in Section \ref{sec:model}.
The resulting relations between the effective velocity dispersion derived from weak lensing and the stellar mass are in Section \ref{sec:vdispscaling}. In
Section \ref{sec:vdispscaling} we demonstrate the one-to-one correspondence between the lensing and spectroscopic velocity dispersions.
We highlight the correspondence between the observational results and the predictions of the Illustris-1 simulations. In the 
discussion (Section \ref{sec:discuss}) we outline subtle issues and potential limitation in the analysis and interpretation of the results.
We conclude in Section \ref{sec:conclusion}.
We use WMAP9 Flat $\Lambda$CDM cosmology throughout this paper ($H_{0}=69.3 {\rm km} / {\rm Mpc/s}, \Omega_{m0}=0.286, \Omega_{b0}=0.0463$) \citep{2013ApJS..208...19H}.

\section{The Data}\label{sec:data}

Our goal is to compare the spectroscopic central stellar velocity dispersions of quiescent objects with
the effective dispersion derived from the weak lensing signal for these objects acting as gravitational lenses. We base our measurement of the relationship between the spectroscopic and lensing signal on two independent
surveys.
We derive the lensing signal from HSC data covering the F2 field of the Deep Lens Survey
(\citet{2002SPIE.4836...73W}; Section \ref{sec:shapes}). The spectroscopy of the lenses comes from the highly complete SHELS survey for galaxies with $R < 20.6$ (\citealp{2005ApJ...635L.125G,2014ApJS..213...35G,2016ApJS..224...11G}, 
Section \ref{sec:lenses})

\subsection{Shape Catalog} \label{sec:shapes}
We use the HSC-$i$ band imaging data for the DLS F2 field ($9^{h}18^{m}00^{s}, +30\arcdeg 00\arcmin00\arcsec$) \citep{2002SPIE.4836...73W} described in \cite[U16]{2016ApJ...833..156U} for shape measurements.
The original DLS F2 field is a $2\times2$ deg$^2$ region but the HSC pointings extend beyond the original boundaries of the F2 field (see Figure 1 of U16). The exposure  is 240 sec  for each pointing.
The typical seeing for the HSC imaging is in the range 0.5--0.7 arcsec.

The images were processed with the standard reduction packages for the HSC Subaru Strategic Survey Program \citep[HSC SSP;][]{Aihara2018},
the \emph{hscPipe} system \citep{Bosch2018}.
We ran version 3.10.2 to reduce the F2 data.
The \emph{hscPipe} system generates a reduced stacked image by applying a standard reduction scheme for the HSC images,
subtracting bias, trimming overscan regions, applying astrometric and photometric calibration, and solving mosaicing solutions and stacking.

Rather than using the shape catalog in \citet{2016ApJ...833..156U} derived with \emph{lensfit},
we use the catalog produced by \emph{hscPipe}. The \emph{hscPipe} catalog is based on an integrated algorithm that takes  information from the instrumental signature removal process into account.
Furthermore \emph{lensfit} derives shape parameters by performing the fit on a thumbnail image of the galaxy. This procedure sometimes fails and reduces the number of usable galaxies. \emph{hscPipe} recovers some of these  galaxies. To construct a clean object catalog, we follow the HSC SSP approach to produce the shape catalog \citep{2018PASJ...70S..25M}.
Appendix \ref{section:cuts} lists the cuts we make to refine the sample (the lensing cut).
We require objects that are not contaminated by artifacts including, for example, cosmic rays and diffraction spikes.
We include only objects that are not blended with a neighboring object.

The resultant number of measured galaxies is $\sim$10\% larger than in the original \citet{2016ApJ...833..156U} shape catalog.
The updated number is compatible with deep number counts based on Suprime-Cam imaging by \citet{2008ApJS..176....1F}.
They derived the galaxy number counts for five $i$-band Suprime-Cam $32{}^{\prime}\times27{}^{\prime}$ pointings.
The small differences among the counts in these independent fields reflect cosmic variance on the Suprime-Cam scale.

\begin{figure}
    \epsscale{1.2}
    \fig{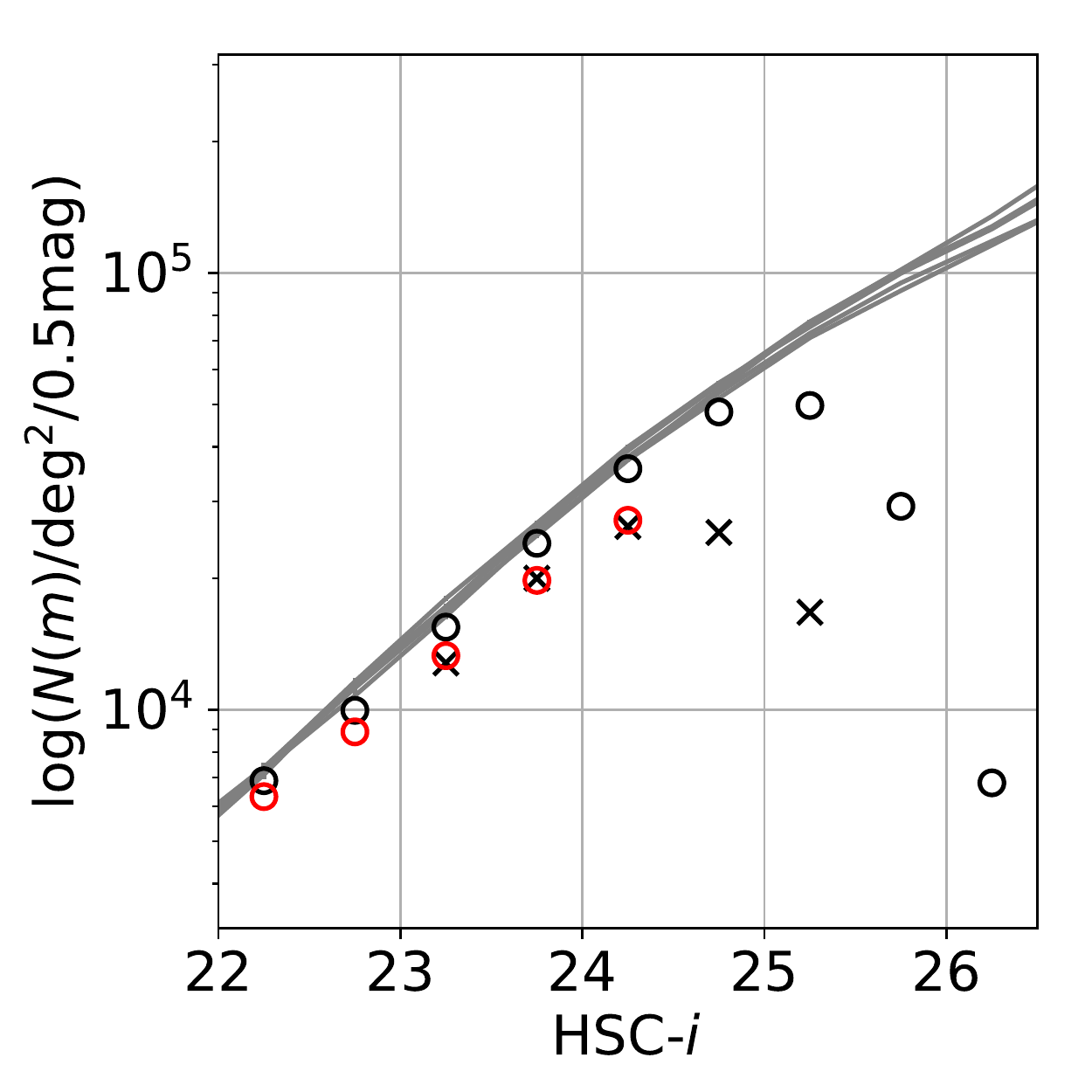}{3in}{}
    \caption{
    Galaxy number density as a function of apparent HSC-i magnitude (Cmodel) for the subsample measured by \emph{hscPipe} for F2 (black open circle) and the subsample used in \emph{lensfit} in \citet[(black crosses)]{2016ApJ...833..156U}. The red open circle represents galaxies survived the lensing cut.
The five curves show deep number counts based on Suprime-Cam imaging \citet{2008ApJS..176....1F}. 
Agreement between the HSC and Suprime-Cam counts is excellent.}\label{fig:numbercount}
\end{figure}

We choose \texttt{"shape.hsm.regauss.*"} measured by the re-Gaussianization technique with an elliptical Laguerre expansion method \citep{2003MNRAS.343..459H} for the shape parameter, $e$.
We then compute the differential tangential surface mass density for each lensing galaxy
\begin{eqnarray}
	\Delta \Sigma(R) = \bar{\Sigma}(<R)-\Sigma(R) = g(R) \Sigma_{\rm cr}
\end{eqnarray}
where the critical surface mass density is
\begin{eqnarray}
	\Sigma_{\rm cr} = \frac{c^2}{4\pi G} \frac{D_{\rm s}}{D_{\rm l} D_{\rm ls}}
\end{eqnarray}
and  $D$ represents the angular diameter distance to the lens (l), source (s) and between the the two(ls).
$g$ is the reduced shear defined as
\begin{eqnarray}
	g = \frac{\langle e \rangle }{2R}
\end{eqnarray}
where $R = 1-0.365^2$ and 
\begin{eqnarray}
	\langle e\rangle = \frac{\sum_i e_i w_i}{\sum_i w_i}, \quad 	{\rm where}\quad w_i = \frac{1}{0.365^2+\sigma_i^2}.
\end{eqnarray}
Once we obtain $\Delta \Sigma(R)$ for each lensing galaxy, we compute an appropriate weighted average for each lensing subsample to construct the stacked lensing signal.

We compute the covariance matrix
\begin{eqnarray}
    &&C_{ij} = \nonumber\\
    &&\left\langle \left( \Delta\Sigma(R_i)-\left\langle\Delta\Sigma(R_i)\right\rangle \right) \left( \Delta\Sigma(R_j)-\left\langle\Delta\Sigma(R_j)\right\rangle \right) \right\rangle,
\end{eqnarray} where $R_i$ is the radius of  the $i$-th bin. For each sample we perform 100 bootstrap resamplings,
In the plots, we take the diagonal component $C_{ii}$ (standard deviation for each bin) as a measure of the error in the stacked lensing shear signal at each radius; we use the full covariance matrix $C_{ij}$ for fitting (Section \ref{sec:model}). 
Use of  the full covariance $C_{ij}$  for fitting improves agreement with previous relations between stellar mass and velocity dispersion (Figure \ref{fig:scaling:sigma-M}).

\subsection{The Lenses}\label{sec:lenses}
The  sample of lensing galaxies is a subset of the Smithonian Hectospec Lensing Survey (SHELS; \cite{2005ApJ...635L.125G, 2014ApJS..213...35G,2016ApJS..224...11G}).
The redshift survey covers the DLS F2 field. 
The redshift survey is 95\% complete to a limiting magnitude $R = 20.6$, where the magnitudes are extrapolated Kron-Cousins $R$-band total magnitudes. Redshifts, stellar masses and $D_n4000$ indices for galaxies in F2 are included in \citet{2014ApJS..213...35G}.

In the full SHELS survey sample, \cite{2016ApJ...832..203Z} derived the central stellar velocity dispersion for 4585 quiescent galaxies. \cite{2016ApJ...832..203Z} review tests of the SHELS velocity dispersion against overlapping objects in the SDSS. They also discuss the small aperture correction.
They select galaxies with $R < 20.6$,  $\log(M_{*}/{\rm M}_{\odot}) > 9.5$ and $D_n4000 > 1.5$ for velocity dispersion measurement.The stellar mass limit ensures that only a small fraction of galaxies have velocity dispersions near the limit set by the Hectospec resolution, $\sim 90~{\rm km s^{-1}}$. More than 85\% of the objects in the total sample have a measured velocity dispersion. The main incompleteness occurs in the redshift range $0.6 < z < 0.7$.

The $D_n4000$ index is the flux ratio between two spectral windows adjacent to the 4000 ~\AA~ break \citep{1999ApJ...527...54B}.
\cite{2010AJ....139.1857W} demonstrate that the $D_n4000$ index is useful for segregating quiescent and star-forming galaxies.
Following \cite{2010AJ....139.1857W}, \cite{2016ApJ...832..203Z} defined galaxies having $D_n4000 \geq 1.5$ quiescent galaxies.
\cite{2016ApJ...832..203Z} measured the line-of-sight (LOS) velocity dispersions $\sigma$
from stellar absorption lines observed through the $1\arcsec.5$ fiber aperture of Hectospec.
They correct the dispersion to a fiducial physical aperture of 3 kpc, approximately  the effective radius of a typical quiescent galaxy (e.g., \citealp{2014ApJ...793...96S}). These aperture corrections are generally small.
The median observational uncertainty in the measured dispersion  $\sigma$ is 38 ${\rm km s^{-1}}$.

\subsection{ The Background Galaxy Redshift Distribution}
We use photometric redshifts to calculate the critical surface mass density. \citet{2013MNRAS.431.2766S} derive a photometric redshift catalog for the original 2 deg$^2$ DLS F2 region. The depth of the photo-$z$ catalog is comparable with the full catalog of sources we detect; thus the 
photo-$z$ catalog can be used to derive the source redshift distribution.

\citet{2013MNRAS.431.2766S} derived photo-$z$s by applying the the BPZ \citep{2000ApJ...536..571B} code to deep $B$, $V$, $R$ and $z$ imaging with 5$\sigma$ limiting magnitudes of 26.0, 26.3, 26.5 and 23.8 mag in 5 $\sigma$, respectively.
The BPZ code provides a best redshift estimate, $z_b$, based on Bayesian priors.
Hereafter we use this $z_b$ as the source photo-$z$s.

The HSC imaging survey extends outside the original DLS survey footprint (See Figure 1 in \citet{2016ApJ...833..156U}). Thus we do not have photo-$z$s for all of the galaxies. To  estimate $\Sigma_{\rm cr}$ throughout the region covered by the HSC imaging, we  simply assume that the background galaxy distribution in the original F2 region applies to the entire survey region. We then estimate $\Sigma_{\rm cr}$ statistically using the  critical surface mass density weighted mean redshift:
\begin{eqnarray}
	\langle z \rangle = \left \langle \left(
    \int_{zl}^{\infty}dz_s (z_s \Sigma_{\rm cr}(z_l, z_s))^{-1} P(z_s) / A_{zl}
    \right )^{-1} \right \rangle_{zl},\label{eq:meanzs}
\end{eqnarray}
where $z_l$ is the redshift of the lensing galaxy, $P(z_s)$ is the normalized background lensed galaxy redshift distribution, and $A_{zl}= \int_{zl}^{\infty}dz_s (\Sigma_{\rm cr}(z_l, z_s))^{-1}P(z_s)$ is a normalization factor.

\begin{figure}
    \epsscale{1.2}
    \plotone{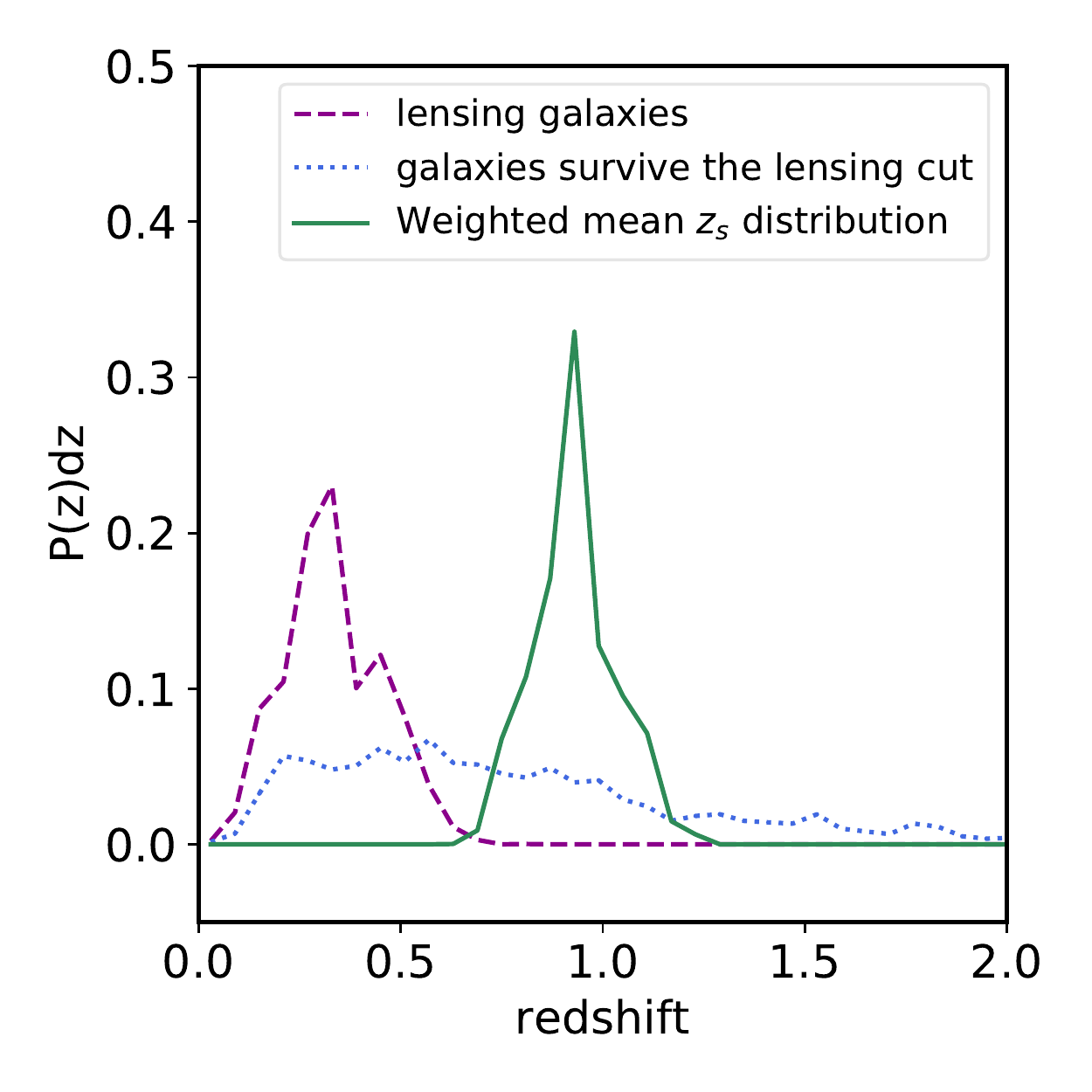}
    \caption{
    Redshift distribution of lensing galaxies (purple dashed), source galaxies that survive the lensing cut (blue dotted). The green solid curve shows the distribution of critical mass density weighted source redshifts.
    }\label{fig:zdist}
\end{figure}

Figure \ref{fig:zdist} displays redshift distributions of the lensing galaxies (purple dashed)  and of the galaxies with photo-$z$s (values of $z_b$) that survive the lensing cut (blue dotted).
The Figure also shows the average of the critical surface mass density weighted mean source redshifts for the set of lenses (within the  bracket in equation \ref{eq:meanzs}) (green solid).
The value of  $\langle z \rangle$ and its dispersion are $0.93\pm0.10$, corresponding to a relative error in the surface critical mass density $\Sigma_{\rm cr}$ in the range  $-0.04<\delta \Sigma_{\rm cr}/\Sigma_{\rm cr}<+0.06$.  

The normalization factor $A_{zl}$ is equivalent to the critical surface mass density for each galaxy.

\subsection{Model}\label{sec:model}

We follow the procedure used for  clusters of galaxies \citep{2011PhRvD..83b3008O} to construct a stacked lensing model for individual galaxies. To represent the matter distribution in the lens, we use an SIS (Singular Isothermal Sphere) rather than NFW (Navarro, Frenk, \& White, \citealp{1997ApJ...490..493N}) profile although some previous lensing studies of ensembles of individual galaxies show that the NFW profile is a better model. 

Our data only allow determination of the profile on scales $ > 100$ kpc. Figure 8 of \citet{2006MNRAS.372..758M} shows that on these scales, the SIS and NFW profiles are indistinguishable within the uncertainties in our measurements. 
This scale  coincides with the range where the  dark matter halo of a galaxy dominates the lensing signal \citep{2013AA...549A...7V}.

Because the Subaru images are deep, light from the extended galaxy halo  and associated structure in the outskirts of galaxies at radii up to 100 kpc
corrupt the lensing signal. We note that this radius is significantly larger than the typical $r_{e}$ for the F2 field \citep{Damjanov2019}.
The depth of the HSC images make it difficult to measure the shapes of the background galaxies projected within 100 kpc of the lens centers. Thus we limit the minimum separation between the center of the lens  and the source to 100 kpc.

Because we compare the lensing signal with spectroscopic velocity dispersions, the SIS model is a direct, simple way to quantify the lensing profile. We employ Equation \ref{eq:model} to quantify the stacked lensing signal:
\begin{eqnarray}
	\Delta \Sigma(R) &=& (1+b)\left[\Delta\Sigma^{\rm SIS}(R)+\Delta\Sigma_{\rm off}^{\rm SIS}(R)\right]\label{eq:model}.
\end{eqnarray}

The first term in the square brackets corresponds to the lensing contribution from the dark matter halo surrounding the central quiescent galaxy. 
\begin{eqnarray}
    \Delta\Sigma^{\rm SIS}(R) = \frac{\sigma_{\rm SIS}^2}{2 G}\frac{1}{R}
\end{eqnarray}
where $\sigma_{\rm SIS}$ is the velocity dispersion for a singular isothermal sphere.

The second term, the one halo term, is a fitting function that represents the complex lensing contribution from massive halos offset from  the position of the lensing galaxy, and/or from foreground/background massive halos within the lensing kernel.
We assume that the offset characterizing this term results from  a random process that obeys a Gaussian distribution with a typical scale of $R_{ \rm off}$.
\begin{eqnarray}
    \Delta\Sigma_{\rm off}^{\rm SIS}(R) = \int  \frac{kdk}{2 \pi} J_2 (k R) \tilde{\kappa}(k) \exp\left( -\frac{1}{2} k^2 R_{\rm off}^2  \right)
\end{eqnarray}
where $\tilde{\kappa}(l)$ is the Fourier transform of an singular isothermal sphere with a velocity dispersion of $\sigma_{\rm ext}$: $2\pi \sigma_{\rm ext}^2/(2 G k)$.

The factor  $(1+b)$ represents a calibration factor that  takes the Photo-$z$ calibration error into account.

\begin{figure*}
\epsscale{1.1}
\plottwo{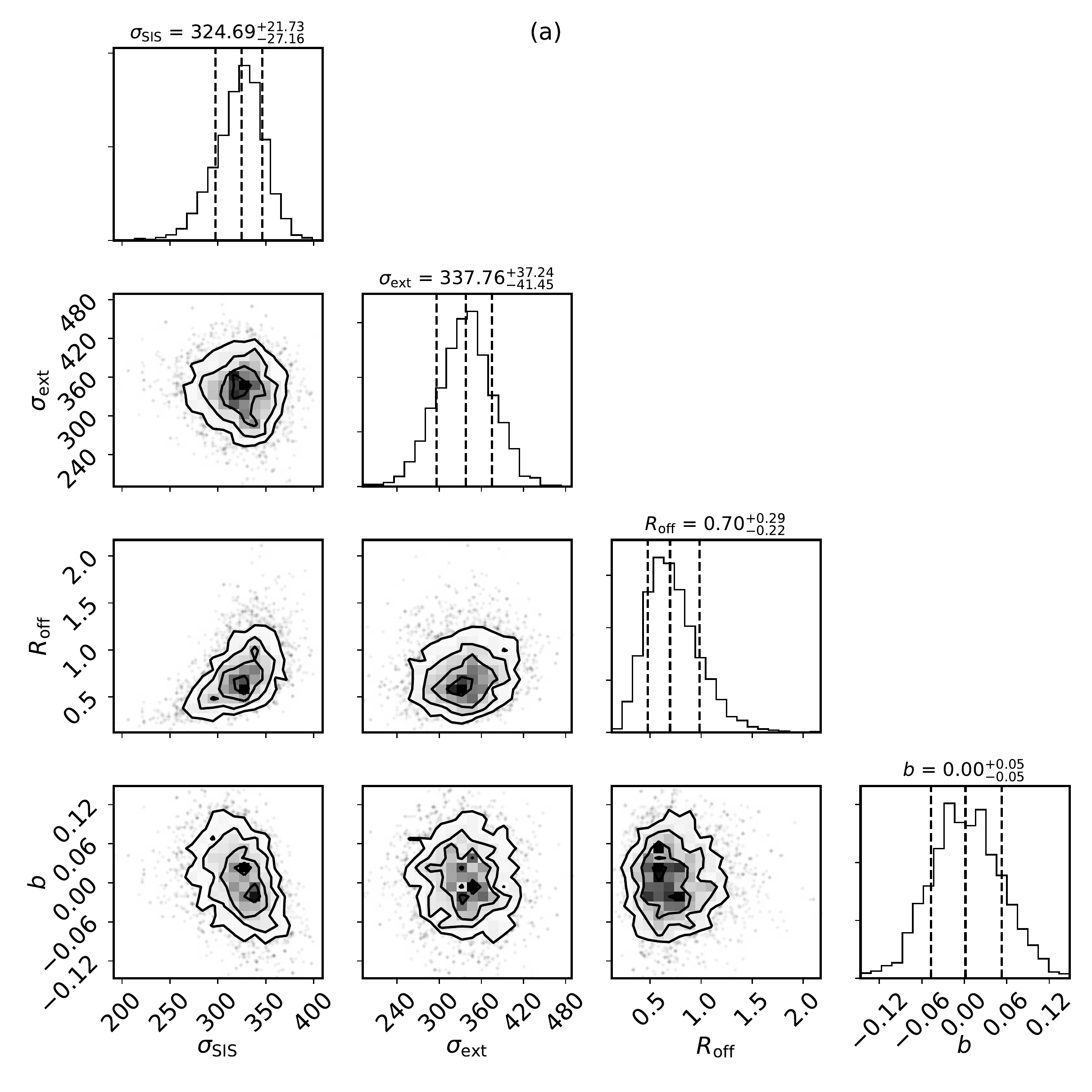}{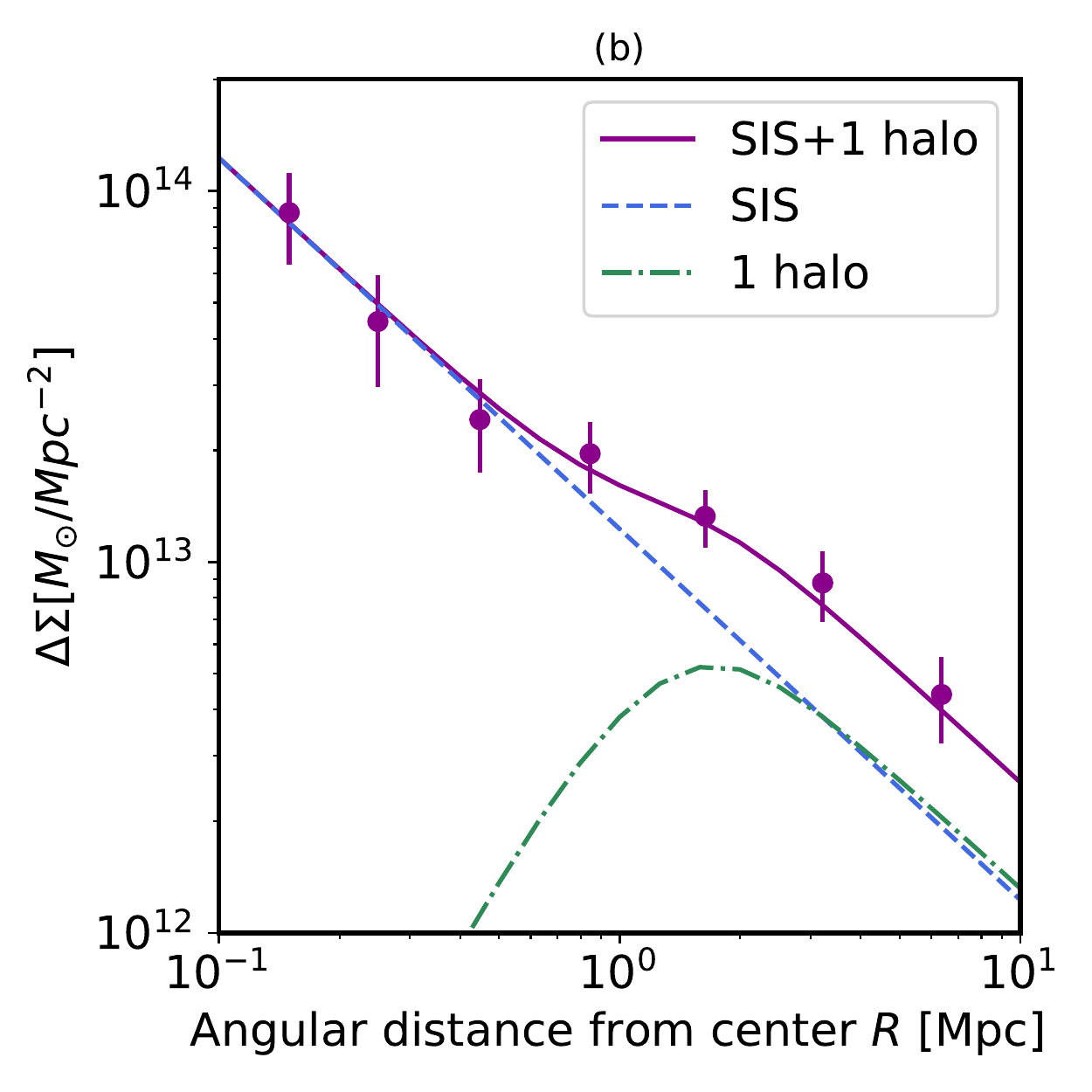}
\caption{
(a) Posterior distribution for model parameters in the most massive velocity dispersion bin.
(b) Fit for the most massive velocity dispersion bin (solid purple). The derived shear, denoted by circles with errors, corresponds to the sum of the SIS (blue dashed curve) and 1 halo terms (green dash-dotted curve).
}\label{fig:SampleFitting}
\end{figure*}

Assuming the 2 model profiles specified by 4 parameters ($\sigma_{\rm SIS}, \sigma_{\rm ext}, R_{\rm off}, b$), we perform 100,000  Markov Chain Monte Carlo resamplings using the Adaptive Metropolis sampler.
We use a uniform distribution for the prior distribution of $\sigma_{\rm SIS}$: $0<(\sigma_{\rm SIS}/{\rm km~s}^{-1})<600$. The other parameters are moderately degenerate. We thus use informative priors: $(\sigma_{\rm ext}/{\rm km~s}^{-1}) \in f(300,50,0,600)$, $(R_{\rm off}/{\rm Mpc}) \in f(1,0.5,0,3)$, $b \in f(0,0.05,-0.15,0.15)$, where $f(\mu,s,a_1,a_2)$ represents a truncated normal distribution with a mean $\mu$ and a standard deviation  $s$ in the range  $(a_1,a_2)$. We determine the prior distributions for $\sigma_{\rm ext}$ and $R_{\rm off}$ from the fact that posterior distributions based on flat priors with the same boundaries are scattered around $\mu, s$. We choose the prior for $b$  to represent the uncertainty in the source distribution. Figure \ref{fig:SampleFitting} (a) shows an example of the posterior distribution. The posterior distributions are nearly indistinguishable from the priors. In other words, there is little constraint on $b$.

Figure \ref{fig:SampleFitting} (b) shows one of the resulting fits to the data. Although our model is simple, the fit overlays the data throughout the range. In other words, the model accounts adequately for the available data.

We also check the level of the B-mode signal, the lensing signal rotated by 45 degree. The B-mode
tests for systematics in the data; ideally there should be no signal in the B-mode. The level of the B-mode signal is consistent with zero within  the statistical errors.

\section{Results}

We begin by describing the properties of the subsamples of the data segregated by velocity dispersion and stellar mass (Section \ref{sec:lensingprofile}). Based on these subsamples we compute the lensing
shear profiles (Section \ref{sec:lensingprofile}). In Section \ref{sec:vdispscaling} we discuss the scaling relations that are the
central result of this investigation.

\subsection{Lensing Profiles}\label{sec:lensingprofile}

We apply the stacked lensing analysis and  model fitting to subsamples of the redshift survey segregated by  central stellar velocity dispersion and by stellar mass. Although stellar mass and velocity dispersion are correlated, the scatter in velocity dispersion at fixed stellar mass is large. Here we define the subsamples we use to measure the lensing signal.

First, we construct a subsample segregated by stellar mass, the stellar mass subsample.
We construct this sample in order to compare our results with  previous relations between
stellar mass ($M_{\star}$) and velocity dispersion ($\sigma_v$) derived independently of weak lensing \citep{2016ApJ...832..203Z}.
We use four stellar mass bins:
$10.0<\log(M_{\star}/{\rm M}_{\odot})<10.5$, 
$10.5<\log(M_{\star}/{\rm M}_{\odot})<11.0$,
$11.0<\log(M_{\star}/{\rm M}_{\odot})<11.5$, and
$11.5<\log(M_{\star}/{\rm M}_{\odot})<12.5$.
To characterize each stellar mass bin, we calculate the mean $\langle M_{\star}\rangle_{\Sigma,cr}$ and standard deviation $\langle (M_{\star}-\langle M_{\star}\rangle)^2\rangle_{\Sigma,cr}$ of the stellar masses both weighted by sigma critical.

Second, we explore the stacked lensing signals based on bins in spectroscopic velocity dispersion, the velocity dispersion subsample.

We divide the lensing galaxies into four bins:
$150 < (\sigma_v / {\rm km s^{-1}}) < 200$,
$200 < (\sigma_v / {\rm km s^{-1}}) < 250$,
$250 < (\sigma_v / {\rm km s^{-1}}) < 300$, and
$300 < (\sigma_v / {\rm km s^{-1}}) < 350$.
To derive a velocity dispersion characteristic of  each subsample, we compute the weighted median velocity dispersions and velocity dispersion as we do for the stellar mass.
Shear signals from subsamples outside of these ranges produce B-mode signals roughly equal to the E-mode signal. We thus do not consider them.
\begin{table}[h!]
\begin{tabular}{ c | c }
\hline
Subsample &  Number of galaxies \\
\hline
\hline
$10.0<\log(M_{\star}/{\rm M}_{\odot})<10.5$    &   798\\
$10.5<\log(M_{\star}/{\rm M}_{\odot})<11.0$    &   2106\\
$11.0<\log(M_{\star}/{\rm M}_{\odot})<11.5$   &   1297\\
$11.5<\log(M_{\star}/{\rm M}_{\odot})<12.5$   &   170\\
\hline
$150{\rm km/s} < \sigma_v < 200{\rm km/s}$  &   1394\\
$200{\rm km/s} < \sigma_v < 250{\rm km/s}$  &   910\\
$250{\rm km/s} < \sigma_v < 300{\rm km/s}$  &   392\\
$300{\rm km/s} < \sigma_v < 350{\rm km/s}$  &   125\\
\hline
\end{tabular}
\caption{Subsamples of lensing galaxies}
\label{table:1}
\end{table}

\begin{figure}
    \epsscale{1.2}
    \plotone{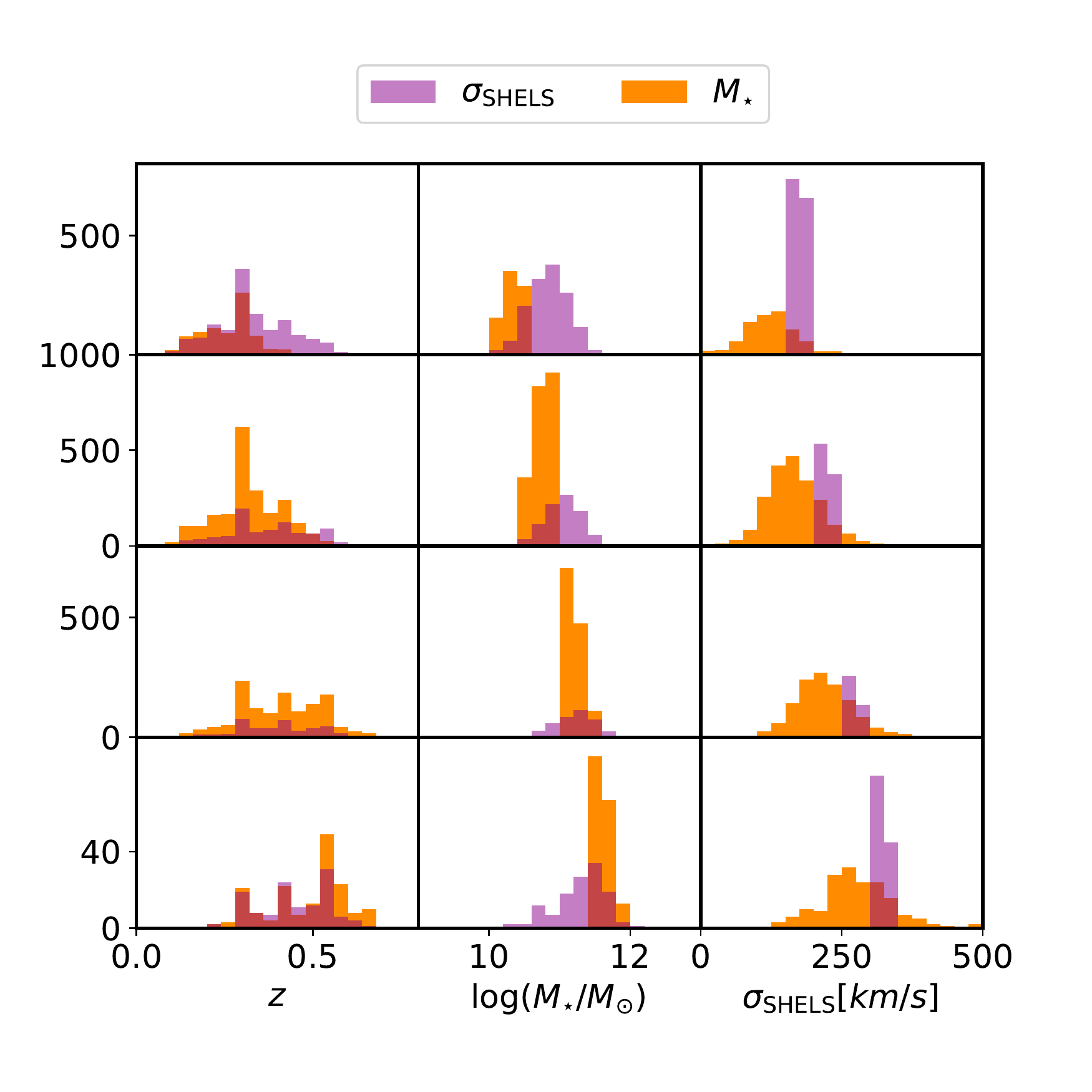}
    \caption{ Distributions of redshift, stellar mass and velocity dispersion for the velocity dispersion (purple) and stellar mass (orange) subsamples. The lowest velocity dispersion (lowest stellar mass) bins are in the top row; the highest velocity dispersion (stellar mass) bins are in the bottom row.  } \label{fig:property}
\end{figure}
Figure \ref{fig:property} shows properties of the subsamples. 
The redshift distribution in the left column shows that the most massive galaxies are present throughout the sample redshift range; the lowest stellar mass objects are only present at lower redshift as expected for an initially magnitude limited redshift survey. 
The middle and the right column each shows clear hard limits depending on whether stellar mass or velocity dispersion is used to segregate the sample. 
At fixed stellar mass the range of central stellar velocity dispersion is large. 

\citet{2016ApJ...832..203Z} discuss the choice of independent variable, $M_{\star}$ or $\sigma$, in the context of selection from a magnitude-limited survey. Depending on the choice, the relation between $\sigma-M_{\star}$ could be affected by observational incompleteness.
Because the observed magnitude is nearly a direct proxy for the stellar mass, the magnitude limit in a narrow range of redshift is a direct proxy for the stellar mass limit (see \citet{Damjanov2019}, Figure 7). The set of measured central stellar velocity dispersions is essentially complete above this stellar mass threshold (Section \ref{sec:lenses} of this paper and references therein).
In contrast,  the stellar mass distribution in a narrow range of $\sigma$ is not complete because the redshift survey is magnitude limited.
When we use the stellar mass as the independent variable, both the stellar mass distribution and the resulting distribution of central stellar velocity dispersion are biased in the same way.
Thus the relation will be free from the incompleteness issue.
For the direct comparison of central stellar and lensing velocity dispersions, these selection issues have no impact on the results.
The results including the mean and a standard deviation may be biased, but both the spectroscopic and lensing samples are
biased in exactly the same way by construction. In other words we compare lensing and spectroscopic velocity dispersions for identical sets of objects. Thus the results are not affected
by the small incompleteness in the spectroscopic survey.

\begin{figure*}
\epsscale{1.1}
\plottwo{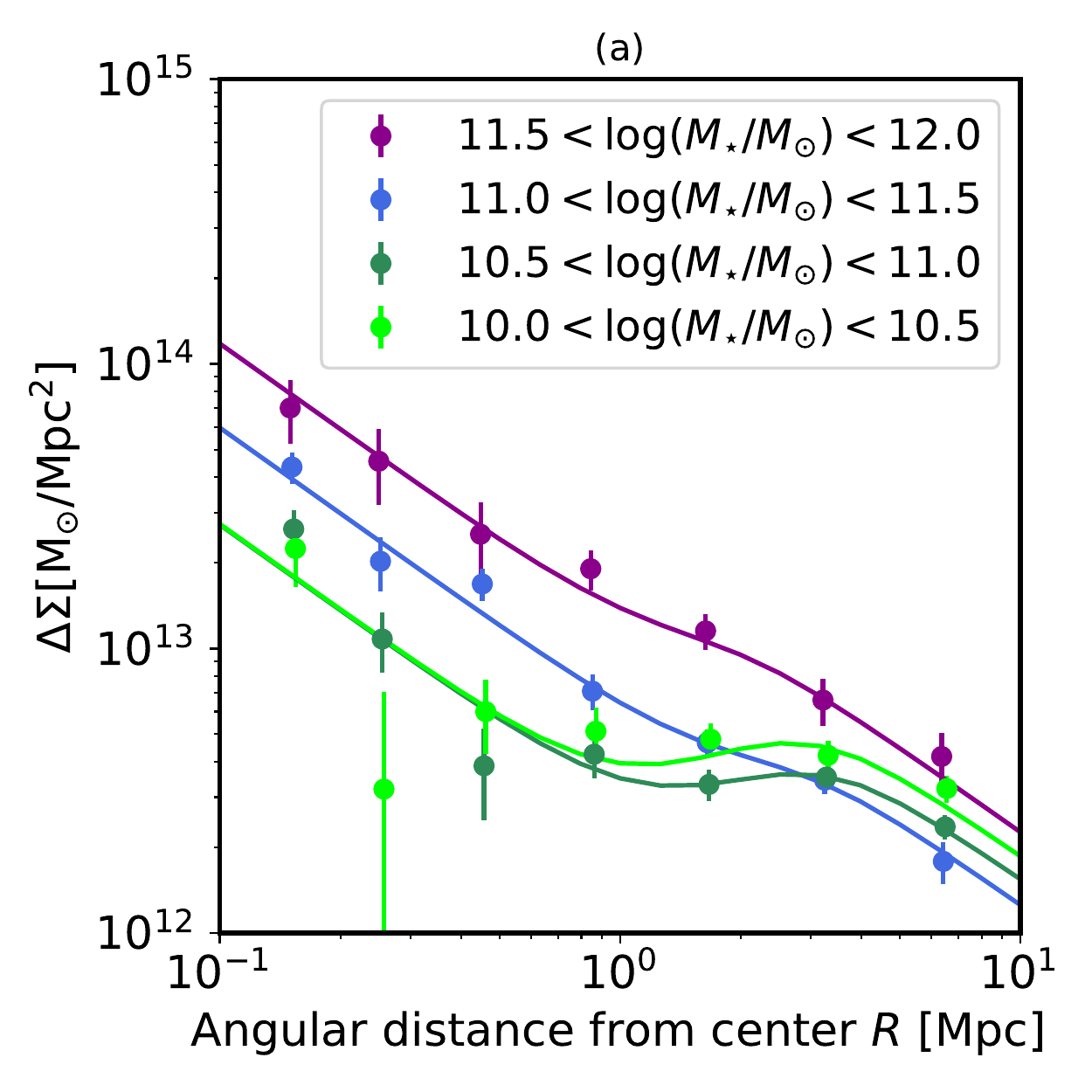}{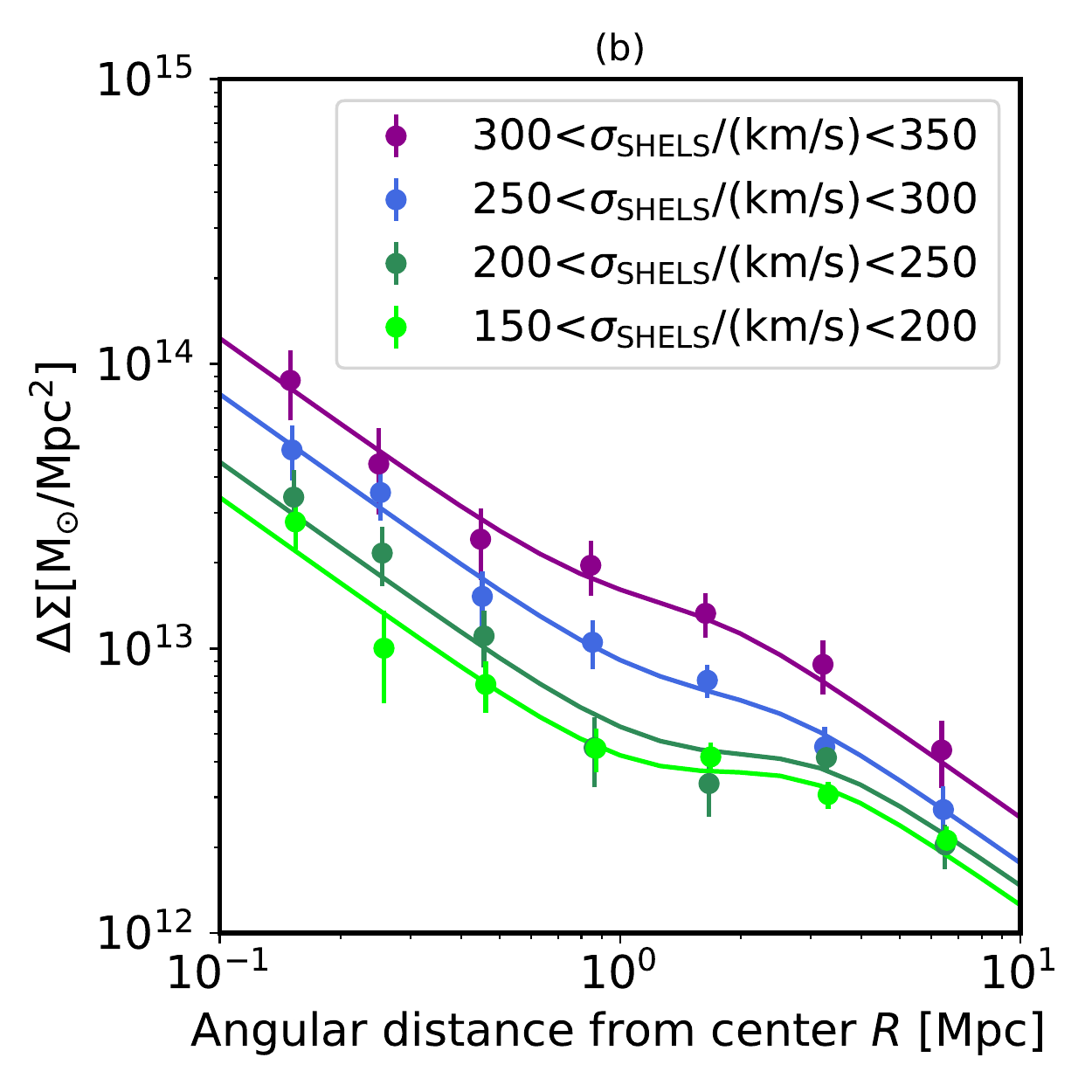}
\caption{
(a) Stacked lensing signal for the lensing galaxy subsamples segregated by  stellar mass. 
(b) Stacked lensing signal for lensing galaxies binned in central stellar velocity dispersion. 
The overlaid curves represent the best fits to the lensing profiles.
}\label{fig:fitresult}
\end{figure*}

Figure \ref{fig:fitresult} shows the stacked lensing shear profiles for  bins  in stellar mass (upper panel) and velocity dispersion (lower panel). The curves show the model fit for each subsample
as indicated in the legend.

The lensing signal is clearly apparent in every case on scales from  0.1 Mpc to 10 Mpc.
Furthermore the model fits provide a good description of  the stacked lensing profiles for the upper two stellar mass bins as well as for the upper two velocity dispersion bins.
For the highest stellar mass bin and for the highest velocity dispersion bin, the inner part of profile fit to the simple SIS model represents the data remarkably well.

In the lowest mass bin the SIS fit is acceptable, but the agreement is not as impressive as for the higher bins in stellar mass and velocity dispersion. In the lowest stellar mass and velocity dispersion  bins,  a bump in the profiles around a few Mpc is especially obvious. The bump is present
but less obvious in the higher stellar mass and higher velocity dispersion bins.
This bump results from the 1 halo term that arises  from miscentering relative to surrounding dark matter halos and/or dark matter halos superimposed in redshift within the lensing kernel (See Section \ref{sec:discuss:1halo} for a discussion of this term). 

\subsection{Velocity dispersion scaling relation }\label{sec:vdispscaling}

\begin{figure}
    \epsscale{1.2}
    \plotone{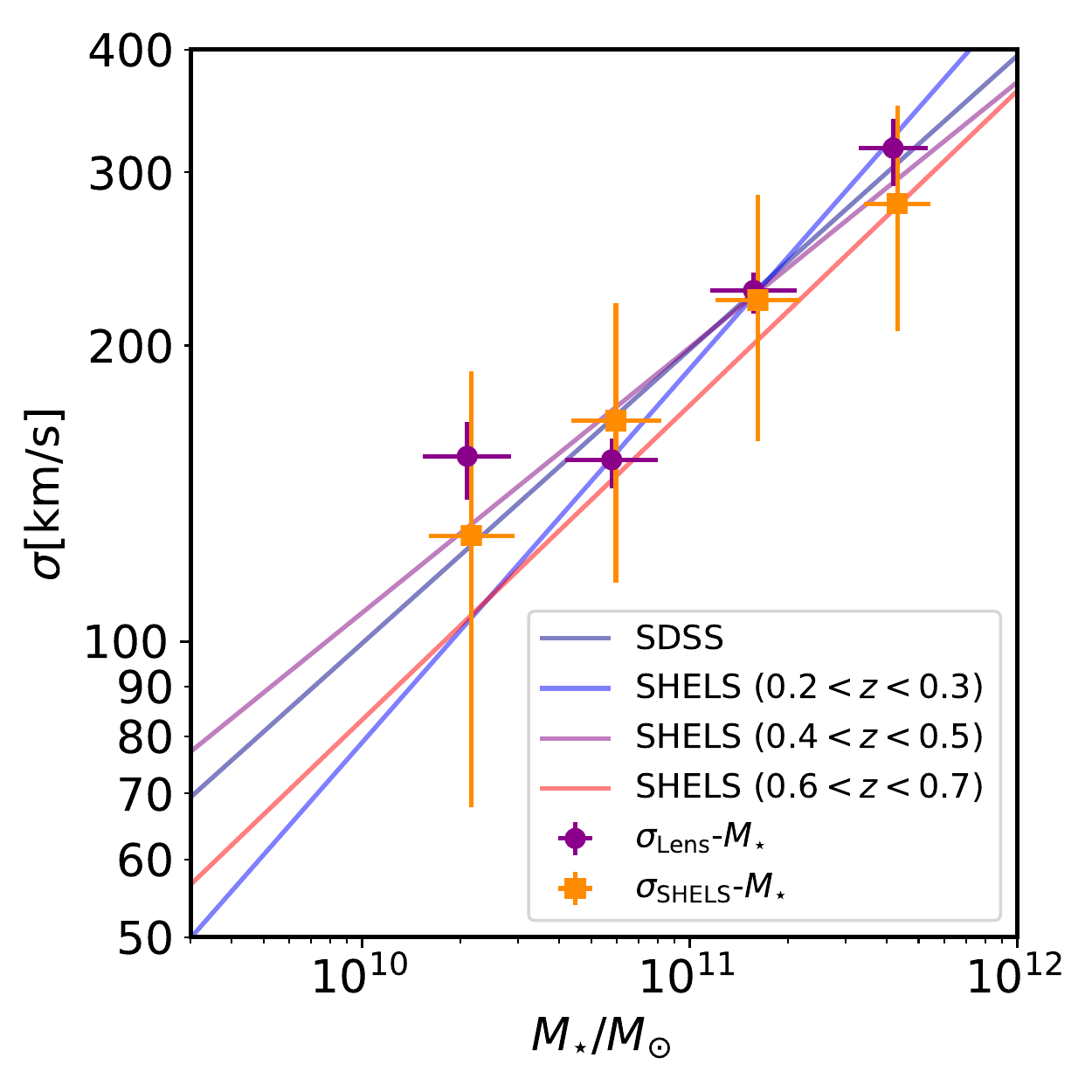}
    \caption{
    Velocity dispersion as a function of stellar mass. Circles represent the lensing velocity dispersion, $\sigma_{\rm Lens}$ on the ordinate; squares represent the central spectroscopic velocity dispersion, $\sigma_{\rm SHELS}$.
    Solid lines show relations for different reshifts from  Table 1 (After "The Full Sample as a Function of Redshift" in \citet{2016ApJ...832..203Z}). From bluish to reddish colors,
    the lines show the result for SDSS, SHELS ($0.2<z<0.3$), SHELS ($0.4<z<0.5$), SHELS ($0.6<z<0.7$).
    The squares are shifted by 1\% for clarity.}\label{fig:scaling:sigma-M}
\end{figure}

We first examine the scaling relation between the lensing velocity dispersion $\sigma_{\rm Lens}$ and stellar mass  $M_{\star}$ (Figure \ref{fig:scaling:sigma-M}). The SHELS sample binned in stellar mass is the basis for this plot. 
The lensing velocity dispersion, $\sigma_{\rm Lens}$ is the median of the posterior distribution (or 50\% percentile) for galaxies in each stellar mass bin; we use the 32\% and 68\% percentile of the posterior distribution to represent the 1 sigma equivalent error.

For comparison, Figure \ref{fig:scaling:sigma-M} also shows the relation between the SHELS spectroscopic velocity dispersion
$\sigma_{\rm SHELS}$ \citep{2016ApJ...832..203Z} and stellar mass once again based on binning the data in stellar mass.
For this comparison, we weight the individual SHELS velocity dispersions by $\Sigma_{\rm cr}$ to mimic the lensing efficiency; we use the standard deviation as the error in the bin. In addition to the overall relation the colored lines show the relations for between SHELS spectroscopic velocity dispersion and stellar mass segregated in redshift from \citet{2016ApJ...832..203Z}.
The bluer color refers to lower redshift; the bluest curve represents a sample from the SDSS.

The slopes and amplitudes of the $\sigma_{\rm Lens}$ and $\sigma_{\rm SHELS}$ versus $M_{\star}$ relations are consistent within the errors. The range of the SHELS relation indicated by the colored curves overlap both the lensing and the spectroscopic results confirming their mutual consistency.

It is worth noting here that the stellar masses
carry a systematic absolute uncertainty of $\sim 0.3~{\rm dex}$ \citep{2016ApJ...832..203Z}.
These systematic issues can affect the amplitude but not the slope of the relation.
Furthermore, although we use photometric redshift to estimate the background distribution statistically,
the estimate remains uncertain because of the limited number of photometric pass bands.
The uncertainty in the mean background redshift of $\pm 0.1$\ produces an uncertainty in the lensing amplitude of $\sim 5$\%.
We take the uncertainty into account and marginalize over when deriving other parameters.
Obviously this systematic error in mean background redshift should not affect slopes of relations between, for example, $\sigma_{\rm Lens}$ and $M_{\star}$.

\begin{figure}
    \epsscale{1.2}
    \plotone{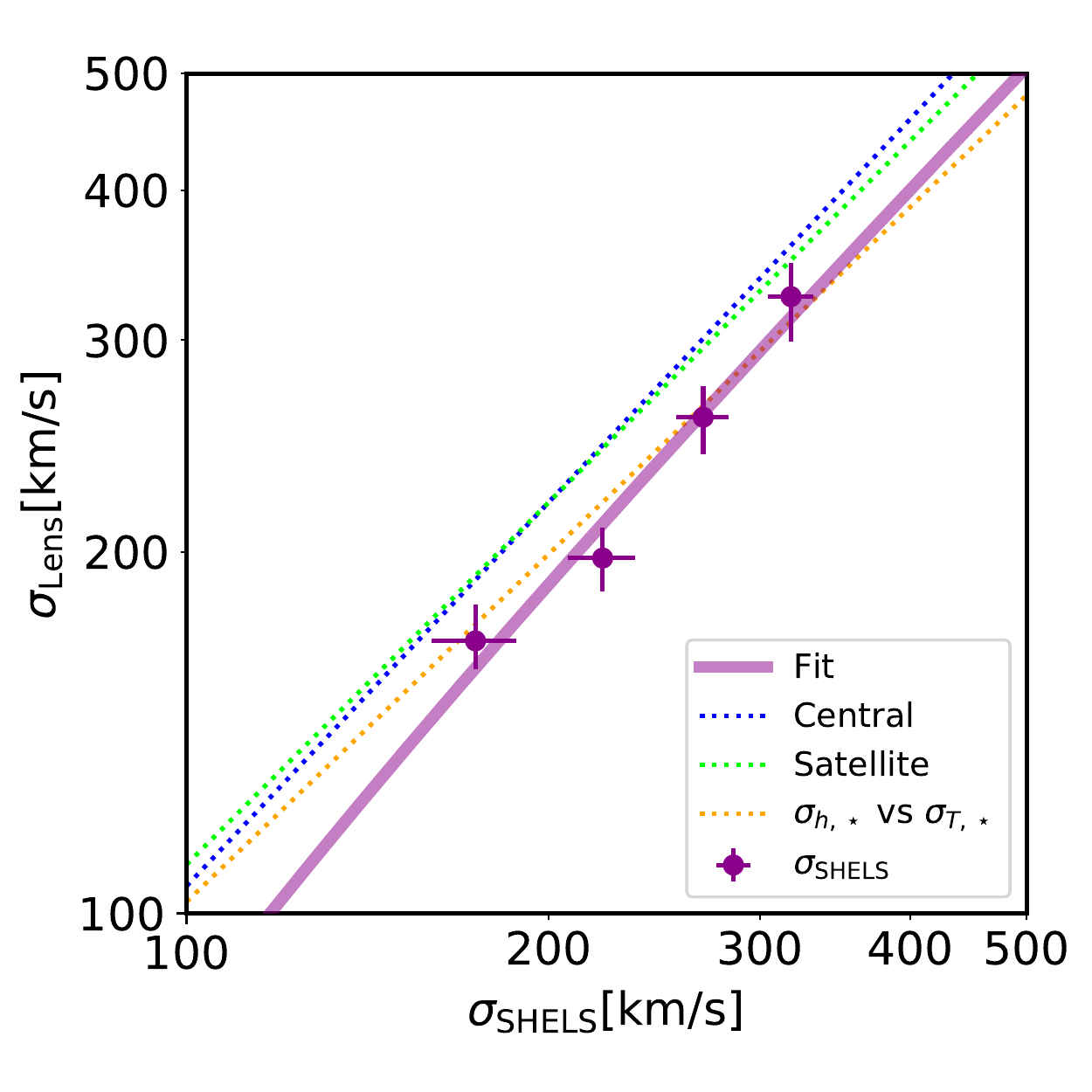}
    \caption{Comparison of lensing and spectroscopic velocity dispersions,
        $\sigma_{\rm Lens}$ versus $\sigma_{\rm SHELS}$. The solid thick line is the best fit linear relation: $\sigma_{\rm Lens} = (1.05\pm0.15)\sigma_{\rm SHELS}+(-21.17\pm35.19)$} in the unit of km/s.
        The dotted lines show relations between the velocity dispersion of the dark matter
        halo as a function of the stellar velocity dispersion for central (blue) and
        satellite (green) galaxies from \citet{Zahid_2018}.
        The (yellow) curve shows the relation between the 3D
        stellar velocity dispersion, $\sigma_{T,*}$, and a simulated line-of-sight measurement, $\sigma_{h,*}$.
        \label{fig:scaling:sigma-sigma}
\end{figure}

Figure \ref{fig:scaling:sigma-sigma} makes the more direct comparison of the lensing derived velocity dispersion,
$\sigma_{\rm Lens}$, and the central spectroscopic velocity dispersion, $\sigma_{\rm SHELS}$.
The lensing velocity dispersions slightly exceed the spectroscopic measurements.

We  fit the points with a line in the unit of ${\rm km}~s^{-1}$:
\begin{equation}
    \sigma_{\rm Lens} = (1.05\pm0.15)\sigma_{\rm SHELS}+(-21.17\pm35.19)
\end{equation}
The derived parameters are consistent with a slope of 1 and an intercept of 0 within errors.
 As in the case of the $\sigma-M_{\star}$ relation, the systematic differences may result in part from error in the estimation of the mean background redshift of the sources. We emphasize again that an error
 in the mean source redshift does not affect  the slope of the relation.

The superimposed curves in Figure \ref{fig:scaling:sigma-sigma} show results from the Illustris-1 simulations \citep{Zahid_2018}.
The dashed curves show relations between the total dark matter halo velocity dispersion and the total stellar velocity dispersion for centrals (blue) and satellites (green).
The red curve denoted by $\sigma_{h,\star}$ vs $\sigma_{T,\star}$ uses the simulations to mimic the data; the relations shows the total stellar velocity dispersion as a function of a line-of-sight proxy that mimics the observations.
Here we note that the simulated relations overlay the lensing results impressively.
We discuss this plot in more detail in Section \ref{sec:discuss:illustris}.

\section {Discussion}\label{sec:discuss}

There are several observational and theoretical issues that affect both the derivation and the interpretation of the scaling relations in Section \ref{sec:vdispscaling}. Here we review and discuss these issues to place the lensing results in a broader context.

We begin by reviewing the investigation by \citet{2013AA...549A...7V} of the relationship between the central stellar velocity dispersion and weak lensing.
Their approach parallels ours but for a
sample of lenses at lower redshift derived from the SDSS. They address issues common to their work and ours and they reach a similar conclusion about the power of the central stellar velocity dispersion as a proxy for the velocity dispersion of the dark matter halo.

Second we discuss the likely interpretation of 1-halo term or, as we use it here, fitting function \citep{2011PhRvD..83b3008O} 
(Section \ref{sec:discuss:1halo}). Then we discuss the limitations that may arise from two underlying astrophysical issues: (1) evolution in the $\sigma-M_{\rm \star}$ relation with redshift (Section \ref{sec:discuss:sigma-M})
and (2) the impact of the distinction between central and satellite galaxies on the scaling relations (Section \ref{sec:discuss:centalandsatellite}).
Finally we discuss the meaning of $\sigma_{\rm Lens}$ and $\sigma_{\rm SHELS}$ in the context of the Illustris-1 hydrodynamical simulations (Section \ref{sec:discuss:illustris})

\subsection {Comparison Sample Based on SDSS }

Because of  the ubiquitous dark matter halos of galaxies, determination of the masses of galaxies is a continuing challenge.
The spectroscopically derived central stellar velocity dispersion for quiescent galaxies has a long history as a potential mass indicator \citep[e.g.][]{2012ApJ...751L..44W,2015ApJ...800..124B}.
Previous investigators explore the correlations between velocity dispersion and other, perhaps less direct, mass proxies including luminosity \citep{1976ApJ...204..668F} and stellar mass \citep{Taylor2010b}.  \citet{2013AA...549A...7V} explore the more direct relationship between velocity dispersion and the dispersion derived from weak lensing for the largest sample to date.

\citet{2013AA...549A...7V} selected a sample of 4000 bulge-dominated (presumably largely quiescent) lensing galaxies at redshift $z < 0.2$ from the SDSS. They
derived the lensing signal from  RCS2 \citep{Gilbank2011} photometry. They take their
central stellar velocity dispersions, $\sigma_{vU}$, from the SDSS DR 7 \citep{Abazajian2008}.

In contrast with \citet{2013AA...549A...7V} our selection of lensing galaxies is based on the spectroscopic parameter D$_n$4000
rather than photometric parameters. Our sample is at a median redshift $z= 0.32$. Thus the samples of lenses are complementary. The HSC photometry provides a source density of $\sim 20$ arcmin$^{-2}$ in contrast with the source density $\sim 6.7$ arcmin$^{-2}$ derived from the RCS2.

Our approach is similar to the analyis of \citet{2013AA...549A...7V} in many ways. They also use an SIS model to interpret the lensing signal and they fit the signal
over a range of radii, 50 kpc to 1 Mpc, 
similar to the range we probe. As they emphasize, the dark matter halo dominates in this radial range. 

\citet{2006MNRAS.372..758M} (in their Figure 8) show that an isothermal profile is indistinguishable from an NFW profile  over the range of scales we consider.
A joint strong- and weak-lensing analysis also shows that the average total mass density profile is consistent with isothermal  over two decades in radius (3-300 $h^{-1}$ kpc, approximately 1-100 effective radii, \citealp{2007ApJ...667..176G}). 

\citet{2013AA...549A...7V} fit a relation $\sigma_{\rm Lens} = a \sigma_{\rm vU} + b$ and find coefficients: 
$a = 0.88 \pm 0.21$ and $b = -5 \pm 50$ km s$^{-1}$ (displayed in their Figure 4),
completely consistent with our result in Figure 7 ($a = 1.05\pm0.15$ and $b = -21.17\pm35.19$ km s$^{-1}$).
\citet{2013AA...549A...7V} conclude as we do that the stellar velocity dispersion, $\sigma_{\rm vU}$ corresponds remarkably well to the velocity dispersion returned by the lensing analysis,
$\sigma_{\rm Lens}$.

\subsection{The 1-halo Fitting Function}\label{sec:discuss:1halo}
\begin{figure}
    \epsscale{1.2}
    \plotone{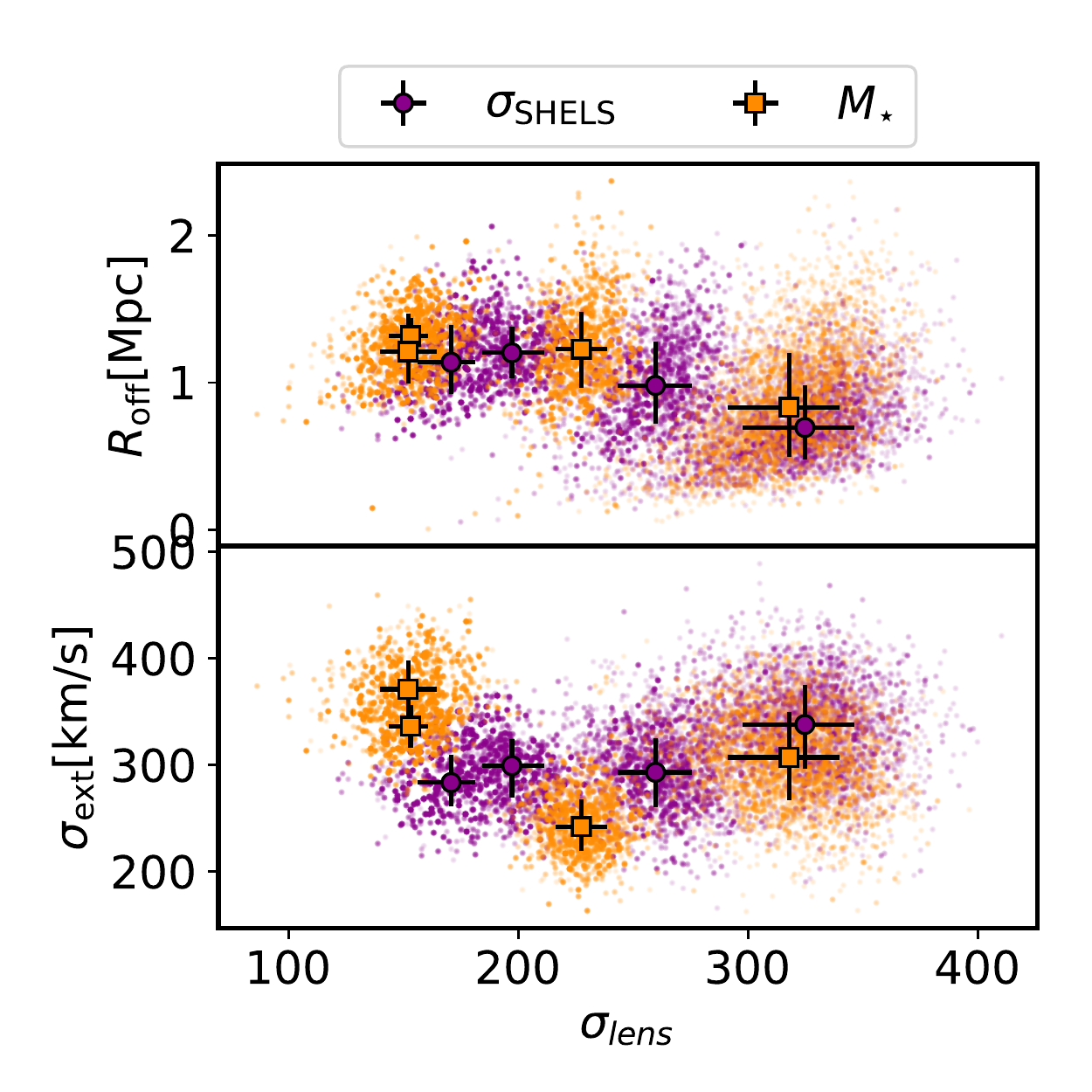}
    \caption{
        Parameters for the 1-halo term. MCMC realizations and aggregated numbers are displayed. We derive the 1-halo fitting parameters by taking the median along with the 16\% and 84\% ranges as the 1-sigma error in  2D space.}\label{fig:parameters}
\end{figure}

We estimate two additional parameters during the fitting procedure $\sigma_{\rm ext}$ and $R_{\rm off}$ that  characterize the 1-halo term. In the original approach where the analysis is applied to the offset between the weak lensing center of a cluster and its apparent optical or X-ray center \citep{2011PhRvD..83b3008O}, the two parameters of this 1-halo term represent, respectively, the velocity dispersion of the cluster and the offset between the lensing and its counterpart measured with other techniques.

When we apply this formalism to individual galaxy lenses, these terms  measure, respectively, the typical velocity dispersion and offset from  surrounding and/or projected dark matter halos. In general these nearby or superimposed halos are groups. In the original application to lensing by a massive cluster, the issue of superimposed halos along the line of sight can safely be ignored because the massive cluster generally dominates the lensing signal. In the case of individual galaxy lenses, several more massive systems, whether nearby or superimposed, can contribute to the lensing profile. Thus in the galaxy-galaxy lensing case the 1-halo term serves as a convenient fitting function with a less direct physical interpretation than in the massive cluster case.

This 1-halo term is responsible for the
bumps at large scale in Figure \ref{fig:fitresult}. Not surprisingly the term is more important for galaxies with either low central stellar velocity dispersion or low stellar mass.

We derive the 1-halo fitting parameters by taking the median along with the 16\% and 84\% ranges as the 1-sigma error in  2D space.
Figure \ref{fig:parameters}  shows these parameters as a function of $\sigma_{\rm Lens}$.
Neither $\sigma_{\rm ext}$ nor $R_{\rm off}$  changes significantly as a function of $\sigma_{\rm Lens}$, although statistical errors are large.

The values of both $\sigma_{\rm ext}$ and $R_{\rm off}$ may be significantly affected by cosmic variance.
The F2 field contains a large number of massive clusters \citep[e.g.][]{2014ApJ...786...93U} at redshifts between 0.3 and 0.4 where the satellites are detectable  and thus probably contribute significantly to these parameters. 
The resulting parameters in the 1-halo term could  differ for a region containing fewer massive systems at redshifts where they can contribute significantly to the lensing signal.

The parameter $\sigma_{\rm ext}$ is related to the satellite fraction $f$ \citep[e.g.,][]{2006MNRAS.368..715M}.
The conversion is: $1/f = ( \sigma_{\rm Lens}/\sigma_{\rm ext})^2 +1$.
The nearly constant $\sigma_{\rm ext}$ as a function of $\sigma_{\rm Lens}$ that we measure implies
a decrease  in $f$ from 0.8 to 0.5 as $\sigma_{\rm Lens}$
increases. \citet[][]{2006MNRAS.368..715M} show that the
satellite fraction decreases with increasing stellar mass and
presumably with the velocity dispersion because it is  correlated with stellar mass.
Thus the behavior of $\sigma_{\rm ext}$ as a function of $\sigma_{\rm Lens}$ is reasonable.
The satellite fractions we derive here exceed the fractions \citet[][]{2006MNRAS.368..715M} derive from a larger sample.
For example $f = 0.2$ for elliptical galaxies in the maximum luminosity bin they consider.
They also demonstrate that the satellite fraction increases with the  local number density (see Figure 8, \citet{2006MNRAS.368..715M}).
Because the number of massive clusters in the F2 field is large \citep[e.g.][]{2014ApJ...786..125S,2016ApJS..224...11G},
it is reasonable that the satellite fraction for F2 is larger. In fact, 
our results are in the range \citet{2006MNRAS.368..715M} obtain at comparable high local densities.

\subsection {Redshift Evolution of the $\sigma$-$M_{\rm \star}$ Relation}\label{sec:discuss:sigma-M}

When we construct the $\sigma$-$M_{\rm \star}$ relation (Section \ref{sec:vdispscaling}), we simply bin the SHELS data in stellar mass and derive the median lensing velocity dispersion, $\sigma_{\rm Lens}$ in each stellar mass bin. Each of the stellar mass bins spans a significant redshift range (See Figure \ref{fig:property}). In this approach, we thus tacitly ignore any redshift evolution in the relation between the central stellar velocity dispersion and stellar mass.

Based on previous work, we can justifiably ignore any redshift dependence in this relation for these massive galaxies ($M_{\rm \star} \gtrsim 10^{10.3} {\rm M}_\odot$) over the redshift range $0.1\lesssim z \lesssim0.7$ explored here.
\citet{2016ApJ...832..203Z} show that for the SHELS data, the redshift evolution of the slope of the $\sigma_{\rm SHELS}-M_{\rm \star}$ relation is consistent with zero. The 
colored dashed lines in Figure \ref{fig:scaling:sigma-M} show the
\citet{2016ApJ...832..203Z} results (from their Table 1) as a function of redshift. Clearly they overlap the lensing results.

Size evolution of the quiescent population complicates interpretation of the $\sigma-M_{\rm \star}$ relation. The traditional fundamental plane encapsulates the
interdependence of the luminosity, velocity dispersion, and size for the quiescent population
\citep[e.g.][]{1987ApJ...313...59D,1987ApJ...313...42D}.
\citet{2009MNRAS.396.1171H} demonstrated that the there is also a fundamental plane relating stellar mass with central stellar velocity dispersion and size.
\citet{2016ApJ...832..203Z} show that ther is no significant evolution in this stellar mass fundamental plane.
Furthermore, \citet{Damjanov2019} show that there is little evolution in size for galaxies with $M_{\rm \star} \gtrsim 10^{11} {\rm M}_\odot$ in the SHELS redshift range. 

At redshifts beyond the range of SHELS, \citet{Belli2014} argue that at redshifts between 0.9 and 1.6,
the smaller sizes of massive galaxies at a fixed stellar mass account for the observed evolution in the relation between stellar mass and velocity dispersion.
At a fixed stellar mass, sizes are smaller and central stellar velocity dispersions are correspondingly larger.
Extending a combined spectroscopic and weak lensing study to greater redshift would enable an understanding of the relationship between the  dark matter halo properties and the evolution observed by \citet{Belli2014}.

\subsection {Central and Satellite Galaxies}\label{sec:discuss:centalandsatellite}

Central galaxies develop in the center of the associated dark matter halo; satellites are generally accreted later and orbit the central. The difference in formation history could lead to a difference in the relationship between the
central line-of-sight stellar velocity dispersion that we measure and the velocity dispersion of the associated dark matter halo \citep[see e.g.][]{Vale2005,Conroy2006,Spindler2017}.

\citet{Zahid_2018}  demonstrate that for central galaxies the Illustris-1 simulated stellar velocity dispersion is an excellent proxy for both the halo velocity dispersion and the halo mass. Figure \ref{fig:scaling:sigma-sigma} shows
their simulated relation between the total stellar and total halo velocity dispersion for centrals  (blue).
We identify the stellar velocity dispersion with $\sigma_{\rm SHELS}$ and the halo velocity dispersion with $\sigma_{\rm Lens}$. The agreement between the observations and the Illustris-1 simulation results is remarkable.

The green line in Figure \ref{fig:scaling:sigma-sigma} shows the Illustris-1
relation from \citet{Zahid_2018} for satellite galaxies. Remarkably, even though satellite galaxy halos are stripped, the total stellar velocity dispersion remains an excellent proxy for the total dark matter halo dispersion. 
The central stellar velocity dispersions of the satellites
are insensitive to the stripping that occurs after their infall, but the masses reflect the stripping history. Thus
the relations between central stellar velocity dispersion and halo velocity dispersion are nearly coincident for centrals and satellites.

Finally, \citet{Zahid_2018} show that when analyzing a magnitude limited redshift survey like SHELS, distinction between central and satellite galaxies is unnecessary. For $M_{\star} > 10^{10.5} {\rm M_{\odot}}$ more than 2/3 of the galaxies in the redshift sample are centrals.
Thus for the massive galaxies we investigate from SHELS, we analyze the sample without making  this distinction.

\subsection {Significance of the Spectroscopic and Lensing Velocity Dispersions }\label{sec:discuss:illustris}

There are at least two fundamental distinctions between the spectroscopic velocity dispersion, $\sigma_{\rm SHELS}$ we measure and the simulated quantities. The blue and green lines in Figure \ref{fig:scaling:sigma-sigma} represent the relation between the total stellar and halo velocity dispersions. They are also measured in 3D and thus naturally take both spatial and velocity anisotropy into account. For lensing velocity dispersion, $\sigma_{\rm Lens}$, we fit the data to an isothermal sphere where the 1D and 3D dispersions are identical. The effective lensing aperture covers a region where the dark matter halo dominates.  In contrast, the spectroscopic observations measure the line-of-sight stellar velocity dispersion projected within a small aperture on the sky.

\citet{Zahid_2018} use the Illustris-1 simulations to explore this issue. They compute the line-of-sight stellar velocity dispersion within the projected half-light radius, $r_e$.
The yellow line in Figure \ref{fig:scaling:sigma-sigma} shows the relationship between this proxy for the observed dispersion and the total stellar velocity dispersion.
The slightly shallower slope of this relation probably reflects a combination of effects including the larger effective radii of more massive galaxies, the dependence of velocity dispersion on aperture, and both geometric and velocity anisotropy.
Because these effects are all secondary, the Illustris-1 results  are completely consistent with
the observed relation between $\sigma_{\rm Lens}$ and $\sigma_{\rm SHELS}$.

\section {Conclusion}\label{sec:conclusion}

A combination of spectroscopy and weak lensing for a large, well-selected samples of galaxies can provide insight into the relation between the galaxies we observe and their ubiquitous dark matter halos.
Here we combine data from two large telescopes, the MMT and Subaru, to compare the central stellar velocity dispersion measured  spectroscopically with the dispersion of the dark matter halo measured by weak lensing.

We derive the basic sample of quiescent galaxy lenses from the SHELS survey \citep{2005ApJ...635L.125G,2014ApJS..213...35G,2016ApJS..224...11G}. This survey provides a set of 4585 lenses with measured line-of-sight central stellar velocity dispersion ($\sigma_{\rm SHELS}$) that is more than 85\% complete for $R < 20.6$, $D_n$4000$> 1.5$ and
$M_{\star} > 10^{9.5}{\rm M}_{\odot}$ \citep{2016ApJ...832..203Z}.
The median redshift of the sample of lenses is 0.32. We measure the stacked lensing signal from HSC deep imaging
\citep{2016ApJ...833..156U}.

We compute the stacked lensing signal for lenses binned separately in stellar mass $M_{\rm \star}$  and central stellar velocity dispersion $\sigma_{\rm SHELS}$.
Following previous work, We use an SIS model with velocity dispersion $\sigma_{\rm Lens}$ to fit the
lensing signal on scales $\gtrsim$ 100 kpc.

We confirm the well-known relation in $\sigma_{\rm Lens}$ vs $M_{\rm \star}$. 
\citet{2016ApJ...832..203Z} determined from $\sigma_{\rm SHELS}$, and it overlaps the  $\sigma-M_{\star}$ relations that span the redshift range of the SHELS survey.

We discuss various issues that could impact the determination of
$\sigma_{\rm Lens}$ including the fitting function, the evolution of the
$\sigma-M_{\rm \star}$ relation, the distinction between central and satellite galaxies, and issues in comparing the spectroscopic 
$\sigma_{\rm SHELS}$ with $\sigma_{\rm Lens}$. We conclude that the relationship between the spectroscopically determined velocity dispersion, $\sigma_{\rm spec}= \sigma_{\rm SHELS}$ and the lensing results, $\sigma_{\rm Lens}$
is insensitive to all of these issues.

The central stellar velocity dispersion $\sigma_{\rm SHELS}$ is directly proportional to the velocity dispersion derived from the lensing $\sigma_{\rm Lens}$.
The independent spectroscopic and weak lensing velocity dispersions probe completely different scales, $\sim3$kpc and  $\gtrsim$ 100 kpc, respectively, and
strongly support the notion that that observable central stellar velocity dispersion for quiescent galaxies is a good proxy for the velocity dispersion of the dark matter halo. 

Our result agrees with and extends the earlier, similar investigation
by \citet{2013AA...549A...7V}. Their  sample of 4000
selected quiescent galaxy lenses is based on photometry rather than spectroscopy. Their lenses are at redshift $ z< 0.2$ rather than a median of 0.32.

The Prime Focus Spectrograph on the Subaru telescope \citep[PFS;][]{2018SPIE10702E..1CT} 
and multi-band deep imaging covering large areas such as HSC SSP \citep{Aihara2018};
Legacy Survey of Space and Time (LSST) will be conducted at the Vera C. Rubin Observatory \citep{2019ApJ...873..111I}, promising an exciting platform for extending this work to much larger samples and to higher redshift.
Larger samples will provide a basis for tests of systematic effects.
These larger samples will also enable exploration of the dependence of the 
relation between the spectroscopically and lensing derived velocity dispersions and  other characteristics of the lenses including stellar mass and size.
Large, deep spectroscopic surveys combined with superb imaging will
enable extension of the comparison of dispersion to greater redshift where they may provide interesting clues to the relative evolution of 
the stellar and dark matter halo components of galaxies.

\acknowledgments{
We thank the referee for helping to strengthen the paper by clarifying several important issues.
We are very grateful to all of the Subaru Telescope and MMT staff.We thank Dr. Okabe for sharing computer resources necessary for reduction and analysis,
Dr. Hironao Miyatake for comments on checking lensing signal,
Dr. Chiaki Hikage for providing comments on constructing the fitting model and,
Dr. Scott Kenyon for comments on effective presentation of the results.

YU was supported by the U.S. Department of Energy under contract number DE-AC02-76-SF00515, Grant-in-Aid Kakenhi from JSPS (26800103) and MEXT (24103003), and Hiroshima Astrophysical Science Center at Hiroshima University.
The Smithsonian Institution supports the research of MJG.
A Smithsonian Clay Postdoctoral Fellowship generously supports the research of HJZ.

The Hyper Suprime-Cam (HSC) collaboration includes the astronomical
communities of Japan and Taiwan, and Princeton University.  The HSC
instrumentation and software were developed by the National
Astronomical Observatory of Japan (NAOJ), the Kavli Institute for the
Physics and Mathematics of the Universe (Kavli IPMU), the University
of Tokyo, the High Energy Accelerator Research Organization (KEK), the
Academia Sinica Institute for Astronomy and Astrophysics in Taiwan
(ASIAA), and Princeton University.  Funding was contributed by the
Ministry of Education, Culture, Sports, Science and Technology (MEXT),
the Japan Society for the Promotion of Science (JSPS), 
(Japan Science and Technology Agency (JST),  the Toray Science 
Foundation, NAOJ, Kavli IPMU, KEK, ASIAA,  and Princeton University.

We used software developed for the Large Synoptic Survey Telescope. We thank the LSST Project for making their code available as free software at http://dm.lsstcorp.org \citep{2019ApJ...873..111I,2010SPIE.7740E..15A}.

Funding for SDSS-III has been provided by the Alfred P. Sloan Foundation, the Participating Institutions, the National Science Foundation, and the U.S. Department of Energy Office of Science. The SDSS-III web site is http://www.sdss3.org/.

SDSS-III is managed by the Astrophysical Research Consortium for the Participating Institutions of the SDSS-III Collaboration including the University of Arizona, the Brazilian Participation Group, Brookhaven National Laboratory, Carnegie Mellon University, University of Florida, the French Participation Group, the German Participation Group, Harvard University, the Instituto de Astrofisica de Canarias, the Michigan State/Notre Dame/JINA Participation Group, Johns Hopkins University, Lawrence Berkeley National Laboratory, Max Planck Institute for Astrophysics, Max Planck Institute for Extraterrestrial Physics, New Mexico State University, New York University, Ohio State University, Pennsylvania State University, University of Portsmouth, Princeton University, the Spanish Participation Group, University of Tokyo, University of Utah, Vanderbilt University, University of Virginia, University of Washington, and Yale University.

The authors wish to recognize and acknowledge the very significant cultural role and reverence that the summit of Mauna Kea has always had within the indigenous Hawaiian community. We are most fortunate to have the opportunity to conduct observations from this sacred mountain.
}

\facility{Subaru (HSC), MMT (Hectospec)}

\software{Astropy\citep[][ascl:1304.00]{2013ascl.soft04002G}, corner\citep{corner}}

\bibliographystyle{aasjournal}
\bibliography{references}

\appendix

\section{Lensing cuts}\label{section:cuts}
In the selection of weak lensing sources, we require that the following parameters be zero.
\begin{itemize}
\item \texttt{deblend.nchild} -- Number of children for this object,
\item \texttt{deblend.skipped} -- Deblender skipped this source,
\item \texttt{flags.badcentroid} -- the centroid algorithm used to feed centers to other algorithms failed,
\item \texttt{centroid.sdss.flags} -- the centroid.sdss measurement did not fully succeed,
\item \texttt{flags.pixel.edge} -- source is in region masked EDGE or NO\_DATA,
\item \texttt{flags.pixel.interpolated.center} -- source's center is close to interpolated pixels,
\item \texttt{flags.pixel.saturated.center} -- source's center is close to saturated pixels,
\item \texttt{flags.pixel.cr.center} -- source's center is close to suspected CR pixels,
\item \texttt{flags.pixel.bad} -- source is in region labeled BAD,
\item \texttt{flags.pixel.suspect.center} -- source's center is close to suspect pixels and
\item \texttt{flags.pixel.clipped.any} -- source footprint includes CLIPPED pixels'
\end{itemize}
The measurement algorithm assumes unblended objects. To filter out the blended objects, we use blendedness parameters $\log_{10}(\texttt{blendedness.abs.flux})<-0.375$.

In addition to the basic cuts above, we apply additional cuts to the set of objects identified as galaxies with a flag of \texttt{classification.extendedness}.  We include only measurements with a high signal to noise ratio as indicated by the following cuts: 
\begin{itemize}
\item {\texttt{shape.hsm.regauss.resolution}$>1/3$},
\item {\texttt{flux.kron / flux.kron.err}$>5$}, and
\item {\texttt{flux.cmodel/flux.cmodel.err}$>15$}.
\end{itemize}
We also reject highly elongated galaxies \texttt{$|e| > 2.0^2$}, which are likely to be blended sources.

\end{document}